\documentclass[aps,pra,showpacs,floatfix, twocolumn,superscriptaddress]{revtex4}
\usepackage{graphicx}
\usepackage{bm,amsmath,amssymb}
\usepackage{dcolumn}
\usepackage{mathrsfs}
\usepackage{cases}
\usepackage{color}
 \usepackage{ulem}
 \usepackage{indentfirst}
\newcommand{\be}{\begin{equation}}
\newcommand{\ee}{\end{equation}}
\newcommand{\bea}{\begin{eqnarray}}
\newcommand{\eea}{\end{eqnarray}}
\newcommand{\la}{\langle}
\newcommand{\ra}{\rangle}

\begin{document}
\title{Percolation of  the Site Random-Cluster Model by Monte Carlo Method}

\author{Songsong Wang}
\affiliation{College of Physics and Optoelectronics, Taiyuan University of Technology, Shanxi 030024, China}
\author{Yuan Yang}
\affiliation{College of Materials Science and Engineering, Taiyuan University of Technology, Shanxi 030024, China}
\author{Wanzhou Zhang}
\thanks{zhangwanzhou@tyut.edu.cn}
\affiliation{College of Physics and Optoelectronics, Taiyuan University of Technology, Shanxi 030024, China}
\author{Chengxiang Ding}
\thanks{dingcx@ahut.edu.cn}
\affiliation{Department of Applied Physics, Anhui University of Technology, Maanshan 243002, China}

\begin{abstract}
We propose a site random cluster model by introducing an additional cluster weight in
the partition function of the traditional site percolation.
To simulate the model on a square lattice, we combine the color-assignation and the Swendsen-Wang methods to design a
highly efficient cluster algorithm with a small  critical slowing-down phenomenon.
To verify whether or not
it is consistent with the bond random cluster model, we measure
several quantities such as the wrapping probability $R_e$,  the percolating cluster density $P_\infty$, and
the magnetic susceptibility per site $\chi_p$  as well as two
exponents such as the thermal exponent $y_t$ and the fractal dimension $y_h$ of the largest  percolating cluster.
We find that
for different exponents of cluster weight $q=1.5$, $2$, $2.5$, $3$, $3.5$ and $4$,
the  numerical estimation of the  exponents $y_t$ and $y_h$ are consistent with the theoretical values.
The universalities of the site random cluster model and the bond random cluster model are completely identical.
For larger values of $q$, we find  obvious signatures of the first-order percolation transition by the histograms and
the hysteresis loops of  percolating cluster density and the  energy per site.
Our results are helpful for the understanding of the percolation of
traditional statistical models.
\end{abstract}
\maketitle

\section{Introduction}
\label{sec:intro}
Broadbent and Hammersley initially presented the concept of
percolation\cite{broad, hamer, Grimmett}, and then Stauffer  introduced
the properties of percolation in detail\cite{stauffer}.
There have been broad applications of percolation:
e.g. fluids in porous medium\cite{pos} ,
the spread of infectious diseases  on complex networks\cite{dis},
the Hall effect with quantum spin \cite{hall},
network vulnerability\cite{neta,netb}, forest fires\cite{fire}, number theory\cite{vardi}, etc $\ldots$

The most studied percolation mo\-dels are percolations on regular  lattices, in which
a site (bond) on the lattice could be occupied (vacant) with probability $p$ (or $1-p$).
At a given critical probability $p_c$, at  least one large cluster, formed by the occupied sites (bonds),
spans to the  opposite boundaries in the lattices\cite{broad, hamer, Grimmett} .

The  construction  of   a site percolation or bond percolation is similar.  However,
they are independent in some respects. For example,
the site percolation transition on the square lattice occurs at
$p_c=0.59274621(13)$ according to the high precision Monte Carlo method\cite{ziff1}, while,
the exact solution indicates that the bond percolation transition point  $p_c=\frac{1}{2}$  on the square  lattice\cite{bondperco}.
In the Monte Carlo simulations near $p_c$,
the configurations  are completely disordered  and the  local structures
in the configurations vary in a significant random
fashion\cite{monte}.

The  invariances behind the configurations are the critical exponents and the universalities,
which are the same for the two types of percolations, without consideration of
the site, the bond, or other microscopic details\cite{universal}.

 Universality connects the phase transitions in a number of lattice statistical models
to the percolation transition.
One important model, the bond random cluster (BRC) model\cite{rcm} created
by Fortuin and Kasteleyn\cite{PWK} in the 1960s, gives us a unified
description of several classical statistical mo\-dels, including
the Ising, Potts\cite{FYW}, Ashkin-Teller\cite{ashkinteller}  and the percolation mo\-dels.
This body of work results in the extensions
of the BRC model and many new possible critical behaviors\cite{tri, guo1, deng2}.

An additional cluster weight factor in the partition function is   the significant difference between the
 bond percolation model and the BRC model.
 Inspired by this,
 we propose a new model, the site RC (SRC) model which
is made by combining  the site percolation and the RC model, and
adding a cluster weight factor in the partition function.

To investigate the critical behaviors of the new SRC model,
we design a cluster-updating  Monte Carlo method and simulate the new model.
Many useful quantities are measured, such as the wrapping probability $R_e$, the  percolating cluster density $P_\infty$
and the magnetic susceptibility per site $\chi_p$.
By performing finite size scaling analysis of the above quantities, the very precise
 phase transition points are obtained.
We also calculate the thermal exponent $y_t$, and  the fractal dimension $y_h$
 of the largest  percolating cluster  in such a way as to check that whether or not
  the universalities of the BRC percolation and the SRC percolation
are completely consistent.

The outline of this work is as follows. Sec.~\ref{sec:model} shows a brief review of the
BRC model and shows how we generalize
 the site percolation model to the SRC model.
Sec.~\ref{sec:method}  describes  the algorithm and several sampled quantities in our Monte Carlo simulations.
Numerical results are then presented in Sec.~\ref{sec:result}.
Conclusive comments are made in Sec.~\ref{sec:conc}.

\section{Model}
\label{sec:model}
\subsection{Potts Model and BRC model}
\begin{figure}[b]
\includegraphics[width=0.4\textwidth]{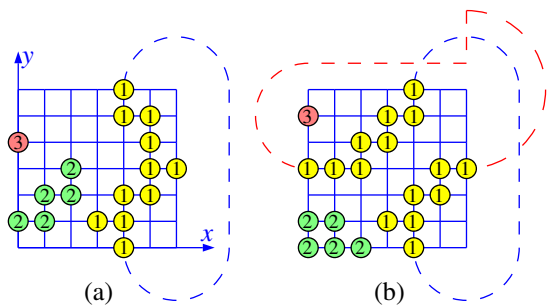}
\caption{(a) A typical configuration of a SRC model on a two dimensional  lattice with size $L=7$,
in which the number of clusters $n_c$ is 3 and the number of the occupied sites $n_s$ is $17$.
The two circles labeled by ``$1$'' in the top and bottom collected by the dashed line, which
 means the first cluster is a wrapping cluster. (b) A cluster labeled by '1', distributed diagonally or "spirally"\cite{ziff1}, wraps around both directions before joining up. }
\label{snap}
\end{figure}
This section provides a brief review of
two classical mo\-dels in statistical physics:
the Potts model\cite{FYW} and  its generalization to the BRC model\cite{PWK}.
The reduced Hamiltonian of the Potts model is:
\begin{equation}
\beta H = -  K \sum_{\langle{ij}\rangle}
\delta_{{\sigma}_{i}, {\sigma}_{j}},
\label{Potts}
\end{equation}
where ${\langle{ij}\rangle}$ means
the nearest-neighbor summation,
 $K$ is the coupling interaction,
$\beta$ is the inverse temperature, $\sigma_i$  is the state variable on the site $i$ and
  can be any natural number less than or equal to $q$.
If $q=2$, the model is identical to the Ising model without an external field, which has two states for each spin.
The partition function of the Potts model is:
\begin{equation}
\begin{aligned}
Z&=\sum_{\sigma} \prod_{\langle ij\rangle}e^{K\delta_{\sigma_i,\sigma_j}}\\
&=\sum_{\sigma} \prod_{\langle ij\rangle}(1 +u \delta_{ \sigma_i,\sigma_j}) ~,
\end{aligned}
\end{equation}
where the symbol $u$ is the   bond  weight and  defined as  $u=e^{K}-1$\cite{sh}.  The above equation can be transformed into:
\begin{equation}
\begin{aligned}
Z&=\sum_{\sigma}\prod_{\langle ij\rangle}\sum_{b_{ij}=0}^{1}(u\delta_{\sigma_i,\sigma_j})^{b_{ij}}\\
&=\sum_{\{b\}}\sum_{\sigma}\prod_{\langle ij\rangle}(u\delta_{\sigma_i,\sigma_j})^{b_{ij}} ~,
\label{zbrc}
\end{aligned}
\end{equation}
where the bond variable $b_{ij}=0$  if $\sigma_i \ne \sigma_j$ while
 $b_{ij}=1$ if $\sigma_i = \sigma_j$.
Through the summation over the spin variable $\sigma$, the partition function Eq.~(\ref{zbrc}) becomes

\begin{equation}
Z_{\text {BRC}} =\sum_{\{b\}} u^{n_b} q^{n_c} ~,
\label{Zrc}
\end{equation}
where the sum is over all bond configurations $\{b\}$, $n_b=\sum b_{ij} $ is the bond number in the configurations,
and $n_c$ is the number of clusters.
The discrete number $q$  now appears as a continuous variable.
Thus,  the BRC model can be regarded as a generalization of the
Potts model. In the limit $q \to 1$, it reduces to the bond-percolation model, whose partition function is:
\be Z= \sum_{\{b\}}(e^K-1)^{n_b} \ee
This form can be easily transformed into:
\begin{equation}
     Z=\sum_{\{b\}}p^{n_b}_b(1-p_b)^{N_b-n_b}
     \label{bpc}
\end{equation}
where $p_b=   \frac{u}{1+u}$ and  $N_b$ is the total number of bonds in the lattice.
The significant difference between the partition functions
of the bond percolation model and the RC model  is that Eq.~(\ref{Zrc}) has  the cluster weight $q^{n_c}$  while Eq.~(\ref{bpc})  does not.
\subsection{SRC model}
Now, we generalize the site percolation to the SRC model\cite{rcm}.
The partition function of the site percolation is:
\begin{equation}
     Z=\sum_{\{\sigma\}}p^{n_s}_s(1-p_s)^{N-n_s},
\end{equation}
where $N=L\times L$ is the total number of sites.
We directly generalize it by introducing a cluster  weight  $q^{n_c}$,
and then derive the partition function of the SRC model as:
\begin{equation}
\begin{aligned}
     Z_{\text{SRC}}&=\sum_{\{\sigma\}}p^{n_s}_s(1-p_s)^{N-n_s}q^{n_c}\\
                &\propto\sum_{\{\sigma\}}u^{n_s}q^{n_c}
                \end{aligned}
\end{equation}
where $p_s=   \frac{u}{1+u}$, $n_s$ is the number of occupied sites, $N-n_s$
is the number of vacant sites, and $p_s$ is the occupation probability for the sites in the
configuration.
The weight of a configuration is given by:
\begin{equation}
W=p_s^{n_s}(1-p_s)^{N-n_s}q^{n_c}
\label{weight}
\end{equation}
As shown in Fig.~\ref{snap}, the weight of the typical configuration is $p_s^{17}(1-p_s)^{32}q^3$.

\section{algorithm and the Sampled Quantities}
\label{sec:method}
\subsection{algorithm}
There are a few efficient methods\cite{md1} to simulate the RC model.
In the present paper, we combine the color-assignation\cite{color1, color2}
and  the Swendsen-Wang\cite{swendsen} methods together to design a
highly efficient cluster algorithm with a small critical slowing-down phenomenon.
Similar methods have been applied  in several papers\cite{ding1,ding2}.
The algorithm  to simulate this model is as follows:

\begin{enumerate}
\item
Initially, all sites are ¡°active.¡±
\item
 Active sites are randomly assigned to be ¡°occupied,¡± with probability $p$ or ¡°vacant¡± with probability $1-
p$. After all sites have been assigned, they are grouped into clusters: if nearest neighbor sites are both
occupied, they belong to the same cluster. Vacant sites don't belong to any cluster.
\item  With probability $1-\frac{1}{q}$, clusters are declared ¡°inactive.¡± The boundary sites-the nearest neighbors
of the sites belonging to an inactive occupied cluster-are also inactive. All other sites are declared
active, in effect erasing their contents.
\item If there are any active sites, return to step 2. Otherwise, we have constructed a configuration that
obeys the statistics of Eq.~(\ref{weight}).
\end{enumerate}

We define  the percolation cluster as follows:
If  any cluster  spans the whole lattice, the configuration is called a percolation configuration.
For a finite system, it can be defined
by various rules. In the present work, a percolation state means there is at least one  ``wrapping'' cluster\cite{wrap}
 in
the lattice and "wrapping" refers to  a cluster that connects itself along  one of the lattice
directions. For example, in Fig.~\ref{snap} (a),
the cluster labeled by "1" is a wrapping cluster, and the wrapping direction is the
vertical direction.
The wrapping cluster is only applicable to a lattice with  periodic boundary conditions.

In Fig.~\ref{snap} (b),
the occupied sites labeled by '1'  are distributed diagonally or "spirally" in the lattice.
In this case, the cluster wraps around both horizontal and vertical directions, which is called the ``single spiral'' configuration\cite{ziff1}.

\subsection{ the  sampled  quantities}
In order to  obtain the critical phase transition points,
we define the wrapping  probability as:
\be
R_e=\la R_x+R_y \ra /2,
\ee
where the subscript $e$ represents
a cluster forming along the $x$ or $y$ direction, and $\la \dots \ra$ denotes ensemble averaging.
If a wrapping cluster exists in the $x$ direction, then  $R_x=1$, otherwise, $R_x=0$.
The rule  is the same for the $y$  direction. If a cluster forming along both $x$ and $y$ direction,
then both $R_x=1$ and $R_y=1$.

The SRC model can be explored in view of site percolation.
Therefore, we can define the order parameter of the percolating cluster density and  magnetic susceptibility per site :
\begin{equation}
     P_{\infty}=\la P \ra =L^{-d}\la n_{\infty} \ra
\end{equation}
\begin{equation}
\chi_p=L^{-2d}\langle \sum_{i=1}^{n_c}n_i^2 \rangle
\end{equation}
where  $n_{\infty}$ is the size (the number of sites) of the  percolating cluster and
$d=2$ is dimensionality of the lattice.
According to the finite-size scaling theory\cite{night,baxter}, the above  parameters provide
us the scaling behavior of them as a function of the system size $L$ and the site occupation probability $p$:
\begin{equation}
\begin{split}
     R_e&=R_e^{(0)}+a_1(p-p_c)L^{y_t}+a_2(p-p_c)^2L^{2y_t}+\cdots\\
    & +b_1L^{y_1}+b_2L^{y_2}+\cdots
    \label{re} ~.
\end{split}
\end{equation}

\[
P_{\infty}=L^{y_h-d}(e_0+e_1(p-p_c)L^{y_t}+e_2(p-p_c)^2L^{2y_t}+\cdots
\]
\begin{equation}
   ~~ \qquad \qquad +f_1L^{y_1}+f_2L^{y_2}+\cdots)
\label{infty}
\end{equation}
\[
\chi_{p}=L^{2y_h-2d}(g_0+g_1(p-p_c)L^{y_t}+g_2(p-p_c)^2L^{2y_t}+\cdots
\]
\begin{equation}
~~ \qquad \qquad +h_1L^{y_1}+h_2L^{y_2}+\cdots)
\label{chi}
\end{equation}

It should be noted that the occupation probability is for the site occupation, instead of the
bond occupation probability\cite{bon}, where $p_c$ is the percolation threshold, $y_t$ is the
 thermal exponent, $y_h$ is the fractal dimension of the percolating cluster, $d$ is the space
 dimension, and $y_1$, $y_2$, $\cdots$, are negative correction-to-scaling exponents.

Eqs.~(\ref{re})-(\ref{chi}) give a model scaling form for various physical quantities.
The three quantities $R_e$, $P_\infty $ and $\chi_p$
are assumed in an analytic function in $p$ and $L$ at the percolation
critical point, so that it has a series expansion here.
These analytic functions, will be used  as a basis
for fitting the numerical data. The three scaling functions that are being
expanded depend on the same scaling variables, but they are, in
general, distinct functions. Hence, when expanded, the expansion
coefficients
 $a_i$, $b_i$, $e_i$, $f_i$, $g_i$, $h_i$ ($i=1,2,\cdots$)  will, in general, be
different. Therefore we use different symbols to denote them.

\subsection{fitting at the critical points}
\label{sec:fitting}

The  fitting functions in   Eqs.~(\ref{infty}) and
(\ref{chi}),  the quantities   $P_\infty$ and
$\chi_p$, depend on the expansion  coefficients.
So it is necessary to   deduce the value of $P_\infty$  and $\chi_p$.
At the percolation point $p_c$, Eqs.~(\ref{infty}) and (\ref{chi}) reduce to:
\begin{subequations}
\begin{align}
 P_{\infty}&=L^{y_h-d}(e_0+f_1L^{y_1}+f_2L^{y_2}+\cdots)\label{infty1}\\
 \chi_{p}&=L^{2y_h-2d}(g_0+h_1L^{y_1}+h_2L^{y_2}+\cdots),
 \label{chip1}
\end{align}
\end{subequations}
which will be used to determine the exponent $y_h$.

To see more readily
the importance of the
corrections to scaling,
we divide  out the leading dependence on $L$ in Eqs.~(\ref{infty1}) and
 (\ref{chip1}) just  using the first two terms. Fitting data according to
\begin{subequations}
\begin{align}
 L^{d-y_h}P_\infty&=e_0+f_1L^{y_1}\label{f1}\\
 L^{2d-2y_h}\chi_p&=g_0+h_1L^{y_1}\label{f2}
\end{align}
\end{subequations}
 will help see clearly  the corrections to the scaling terms.

\section{Results}
\label{sec:result}
Firstly, we do a Monte Carlo simulation of the SRC model on the square lattice with the above algorithm.
We find the algorithm has  a small critical slowing-down phenomena with $q\leq4$
 and consequently we sample between  every two Monte-Carlo steps.
As the system enters into  equilibrium states, we take $10^8$ samples to
calculate each quantity for the system  sizes $8 \leq L \leq 64$, and we take $10^7$
samples  for the system sizes $128\leq L \leq 256$\cite{FDB}.

To obtain the critical point $p_c$,  and  the exponent $y_t$,
we perform  a finite-size scaling analysis  of the wrapping probability $R_e$ for various system sizes
near the critical occupation probability $p_c$. At the critical point $p_c$, we calculate the percolating cluster density $P_\infty$ and  the magnetic susceptibility per site $\chi_p$ to obtain the exponent $y_h$.
We also study the cases for larger values of $q$, such as $q=10$ and find a interesting first-order phase transition.

\label{sec:res}

\subsection{Theoretical and numerical exponents $y_t$ and $y_h$ for q=1.5-4 }
The theoretical values of the exponents  $y_t$ and $y_h$ can be obtained by the Coulomb gas method\cite{coulomb} or conformal invariance\cite{conformal}, and
they are given by:
\begin{subequations}
\begin{align}
\sqrt{q}& =-2 ~  \cos ( \pi g ) ,\\
 y_t&=3-\frac{3}{2g},\\
 y_h&=1+\frac{g}{2}+\frac{3}{8g}.
\end{align}
\end{subequations}
where  the coupling constant $g$ of the Coulomb gas is in the range $1/2 \le g \le 1$.
According to the above equations, the theoretical values of the both exponents will be shown in
the following section.

The numerical results by Monte Carlo method  are listed in table \ref{Tab:table}.
For  $q=1.5,~2,~2.5,~3,~3.5$ and $4$,  the percolation
threshold $p_c$, the wrapping probability $R_e$,  the thermal exponent $y_t$, and the fractal dimension of the percolation
 cluster $y_h$ are obtained in the same way, which will be discussed in detail next subsections.
We find that  for the range $q=1.5-3$, the numerical results  $y_h$ and $y_t$
are very consistent with  the theoretical values.
 For $q=3.5$ and  $4$, the precision of the critical point and  the exponents are
 lower than the case with other values of $q$, due to the logarithmic correction\cite{log1,log2, log3}.

\begin{table}[htb]
\caption{Numerical results(N) for the percolation threshold $p_c$, the wrapping probability $R_e$, the thermal exponent $y_t$,
and the fractal dimension $y_h$ from  $\chi_p$. Theoretical
predictions(T) are included where available by the Coulomb gas method\cite{coulomb} or conformal invariance\cite{conformal} .
The estimated errors in the last decimal place are shown between parentheses.}
  \begin{tabular}[t]{c|l|c c l   l  }
\hline
\hline
q& &
$p_c$ &$R_e$&~~$y_t$ & $y_h \leftarrow\chi_p$\\
\hline
1.5&~~N& ~0.726525(2)~~&0.5822(3)~~&0.884(4)~~& 1.8831(7) \\
&~~T~~&~ --& --&0.887~~&1.8832\\

2&~~N& ~0.805000(1)~~&0.6270(1)~~&1.000(5)~~& 1.8750(5)\\
&~~T& ~--& -- &1.000~~&1.8750\\

2.5&~~N& ~0.854411(2)~~&0.6637(3)~~& 1.101(7)~~& 1.8698(4) \\
&~~T & ~-- &--&1.102~~&1.8697\\

3&~~N& ~0.887435(1)~~&0.6955(2)~~&1.196(5)~~& 1.8664(7) \\
&~~T & ~-- &--&1.200~~&1.8667 \\

3.5&~~N& ~0.910600(2)~~&0.7242(8)~~&1.311(8)~~& 1.867(1) \\
&~~T&~ -- &-- &1.305~~&1.866\\

4&~~N& ~0.927476(1)~~ &0.750(1)~~& 1.44(7)~~&1.88(1) \\
&~~T&~ -- &--&1.50~~&1.88\\
\hline
\hline

\end{tabular}
\label{Tab:table}
\end{table}

\subsection{$q=1.5$, detailed analysis}
\begin{figure}[b]
\vskip 0.5cm
\includegraphics[width=0.45\textwidth]{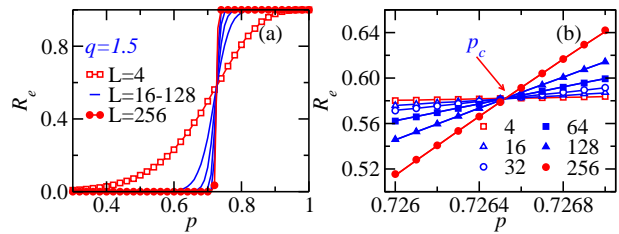}
\caption{Wrapping probability $R_e$ versus site-occupation probability $p$ at  $q=1.5$ in the ranges  (a) $0.2<p<1$ and (b)
 $0.7260<p<0.7272$, with  different  sizes $L=4,~8,~16,~32,~64,~128$, and $256$. The critical point is $p_c=0.726525(2)$ and $R_e=0.5822(3)$.
 The error bars are smaller than the symbols. The lines in the right figure are plotted to guide the reader.}
\label{wrap}
\end{figure}

As shown in Fig. \ref{wrap}(a), we calculate the wrapping probability $R_e$ as a function of  site  occupation
probability  $p$ at $q=1.5$ for lattices with  different  sizes $L=4,~8,~16,~32,~64,~128$, and $256$.
In the limit $p \rightarrow 0$, no sites are occupied and hence no clusters exist and $R_e=0$.
In the limit $p \rightarrow 1$, all sites are occupied and a wrapping cluster forms and $R_e=1$.

In the region of the critical points, i.e.,  $0.7260<p<0.7272$,
the data looks nearly linear  as shown in Fig. \ref{wrap}(b).
 Using the Levenberg-Marquardt least-squares method\cite{lm} and {\relax Eq.}~(\ref{re}), we find that the critical percolation probability is at $p_c=0.726525\text(2\text)$.
Correspondingly, the thermal exponent is $y_t=0.884(4)$, which is consistent with the
 theoretical result  $y_t=0.887$. 

 In the fitting procedure, the chi-square
 \be
 \chi^2=\sum_{L_i}\left(\frac{R_e\left(p,L_i\right)-R_e^{fit}\left(p,L_i\right)}{\sigma^2_i}\right)^2
 \ee
is performed\cite{fit1,fit2} by summing over the sizes $L=16, 32, 64, 128, 256$.
The order of magnitude of chi-square  is 10.
 The ratio  of chi-square to  degree of freedom of fit  $\chi^2/d.o.f$ is  $1.04$, which was
 thought to be a moderately good fit.
$\sigma_i$ is the error of $R_e$ measured by the Monte Carlo method.  $R_e^{fit}$ represents the fitting function of $R_e$ in
 Eq. (\ref{re}).
The results with $L=4, 8$ are dropped and the higher terms in the expansion are also dropped, i.e., $a_i=0,i=3,4,\cdots$ and
$b_i=0, i=2, 3, \cdots$.

\begin{figure}[t]
\includegraphics[width=7.7cm,height=6cm]{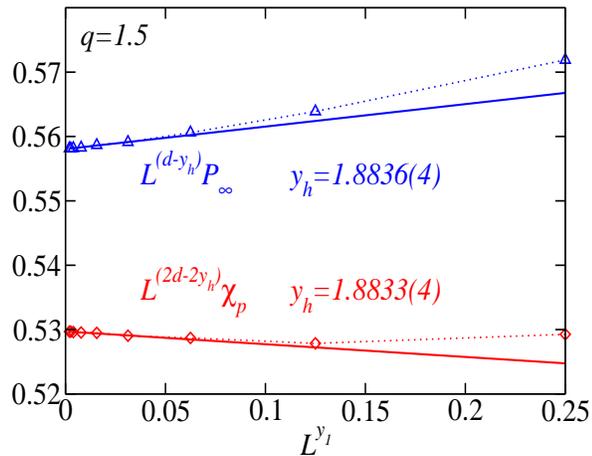}
\caption{  $L^{2d-2y_h}\chi_p$ and  $L^{d-y_h}P_{\infty}$ versus  $L^{y_1}$  of the SRC model for $q=1.5$ on the square lattice with $L=32, 64, 128, 256$ and $512$. The exponent is fixed being $y_1=-1$. The two solid lines
represent fits to the data points according to Eqs.~(\ref{f1}) and
 (\ref{f2}). The fitted exponent $y_h$ are 1.8836(4) and  1.8833(4) from  $L^{d-y_h}P_{\infty}$ and $L^{2d-2y_h}\chi_p$   , respectively.
 The dashed lines are plotted to guide the reader.}
\label{loglog}
\end{figure}

Figure~\ref{loglog} displays
the plot $L^{2d-2y_h}\chi_p$ and  $L^{d-y_h}P_{\infty}$ versus  $L^{y_1}$ at the percolation point.
The plot symbols for systems with sizes $L=32-512$ sit in the fitted lines very well, as expected.
For small systems with sizes  $L=4, 8, 16$, the plot symbols
 deviate from  the fitted line.
  Obviously, the correction-to-scaling of  $L^{d-y_h}P_{\infty}$  is similar
with that of $L^{2d-2y_h}\chi_p$\cite{bon}.
In the real fitting procedure, we neglected  the data with sizes $L=4-16$ and  the order of
magnitude of the residual  equals  to $10^{-9}$, which means the results are still reliable.

 The  leading correction-to-scaling exponent\cite{y1} is known to be $y_1  \approx
-1$. A least-squares criterion was used to fit the data
 with $y_1$ being fixed at $-1$.
 By fitting the data of $L^{d-y_h}P_{\infty}$, the exponent  is  fitted and found to be $y_h=1.8836(4)$.
However, by the fitting  of  $L^{2d-2y_h}\chi_p$, the exponent becomes $y_h=1.8833(4)$, which is consistent with
the result from $P_{\infty}$.
The slopes $f_1=0.035(7)$ and $h_1=-0.020(2)$ for both fitted lines and  the first expanded coefficients
  $e_0=0.5580(5)$ and  $g_0=0.5297(2)$ are also obtained.



For larger systems, the correction terms  in Eqs.~(\ref{f1}) and (\ref{f2}) are far less than the first terms $e_0$ and $g_0$ at the
critical points and therefore the power law $P_{\infty} /\chi_p \propto  L^{d-y_h}$ can be obtained by neglecting
the correction terms.
In fact, scaling theory for percolation
(e.g. see \cite{deGennes,Grimmett,stauffer}) predicts that phase transitions exhibit scaling properties or ``power laws''.
Moreover, power laws like Newton's gravitational law or Coulomb's law or even Lotka's law for publication rates\cite{lotka} are ubiquitous
and it is reassuring to recover a power law here as well.

\begin{figure}[t]
\includegraphics[width=0.45\textwidth]{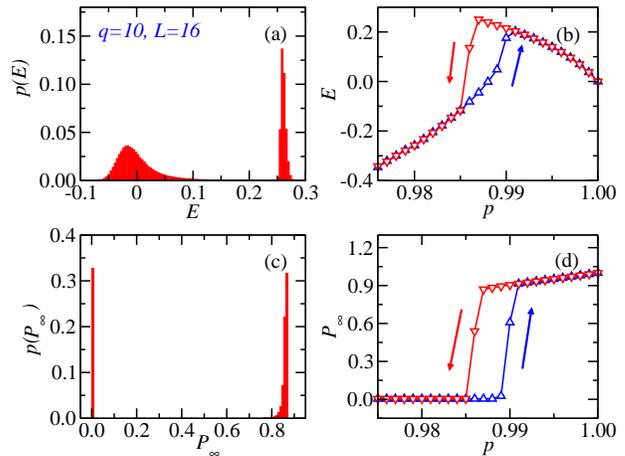}
\caption{The signature of the first-order phase transition for the SRC model  at $q=10$ on a $16 \times 16$ square lattice.
  Histogram of energy per site $E$ (a) and   the percolation  strength  $P_\infty$ (c) at the critical point $p_c=0.987$.
 Hysteresis loop of the both quantities $E$ (b) and $P_\infty$ (d)  around the critical point  $p_c$.}
\label{hist}
\end{figure}

\subsection{$q=10$, a first-order phase transition}

 Figure~\ref{hist} (a) shows a histogram of the energy per site  $E$ at the critical point $p_c=0.987$,
in which  the double distribution  is a typical signature of  the first-order phase transition from the non-percolation phase
to the percolation.  We obtain the histogram  in such a way. Firstly, we initialize a configuration by assigning   each site
with an occupied or  an empty state,
a probability of $1/2$.
 After the system enters into an equilibrium state,  we measure the energy per site $E$.
 We repeat the above steps until the shape of the histogram converges.

To confirm the first-order of the percolation transition, Fig. \ref{hist} (b) shows the hysteresis loop
  around the critical point region, i.e., $0.975<p<1$. The  hysteresis loops have been
    observed both in classical\cite{deng} and quantum systems \cite{steep,hysloop1,hysloop2, hysloop3}.
     To form a closed hysteresis  loop, we start at $p=0.975$. Then
     we increase the occupation probability $p$ and sample the energy per site $E$.
In the simulation, we use the configuration of the previously completed simulation for
a given value of ``$p$'', as the (new) initial configuration of the simulation of another value
of ``$p$''.
  The energy per site $E$ of the system does not jump to a higher value immediately until
        $p$ exceeds over a short distance of  the transition point $p_c$.
After  $p$ reaches $1$, we decrease $p$  in the same way with regards to the initialization of configurations. A closed
      hysteresis loop forms when $p$ becomes smaller than $p_c$.
We repeat similar steps for the  $P\infty$ and the results are  shown in Figs.~\ref{hist} (c) and (d).

\section{Conclusion}
\label{sec:conc}
In conclusion, we have proposed a new statistical model, which
can be considered as a SRC model with an additional
cluster weight in the partition function with respect to the traditional
site percolation model.

We have also designed a color-assigned cluster updating Monte Carlo simulation algorithm
suffering little from the  boring critical slowing-down phenomena.

Both of the BRC and SRC percolation mo\-dels have the same universality
by simulations of the SRC model on the square
lattice and  behaviors of  the quantities $R_e$, $P_{\infty}$, $\chi_p$,
$y_t$ and $y_h$.

At the critical phase transition point the case of $q=1.5$,  the
correction-to-scaling of  $P_\infty$ is close to
 that of $\chi_p$.
The fitted exponent $y_h$ from $P_\infty$ has the same precision with that from $\chi_p$.
For $q=4$, the estimation of exponents $y_t$ and $y_h$ is less precise due to the log-correction.
For $q=10$, the  obvious first-order transition is observed.

Our results can be  considered as a first study of  the
counterpart for the  BRC percolation model
and are helpful for the understanding of the percolation of
traditional statistical  mo\-dels.

\acknowledgments
W. Zhang would like to thank T. C. Scott in helping him prepare this manuscript.
W. Zhang is supported by the NSFC
under Grants No.11305113 and No. 11204204, Foundation of
Taiyuan University of Technology 1205-04020102. C. Ding is supported by the
NSFC under Grant No. 11205005, Anhui Provincial Natural
Science Foundation under Grant No. 1508085QA05 and 1408085MA19.
T. C. Scott
is supported in China by the project GDW201400042 for the
high end foreign experts project.

\end{document}